\documentclass[superscriptaddress,amsmath,amssymb,prl, preprint]{revtex4}
\usepackage{graphicx}
\usepackage{subfigure}
\usepackage{graphicx}
\usepackage{dcolumn}
\usepackage{bm}
\usepackage{amsmath}
\usepackage{setspace}
\usepackage{mathrsfs}

\newcommand{\DvR}[1]{\left| #1 \right\rangle}
\newcommand{\DvL}[1]{\left\langle #1 \right|}

\newcommand{\ident}[0]{\leavevmode\hbox{\small1\normalsize\kern-.33em1}}%

\newcommand{\Sp}[0]{S}
\newcommand{\SpL}[0]{S_L}
\newcommand{\SpR}[0]{S_R}

\begin{document}

\title{Fidelity enhancement by logical qubit encoding}

\author{Michael K. Henry}
\affiliation{Department of Nuclear Science and Engineering, Massachusetts Institute of Technology, Cambridge, Massachusetts 02139, USA}
\author{Chandrasekhar Ramanathan}
\affiliation{Department of Nuclear Science and Engineering, Massachusetts Institute of Technology, Cambridge, Massachusetts 02139, USA}
\author{Jonathan S. Hodges}
\affiliation{Department of Nuclear Science and Engineering, Massachusetts Institute of Technology, Cambridge, Massachusetts 02139, USA}
\author{Colm A. Ryan}
\affiliation{Institute for Quantum Computing, University of Waterloo, Waterloo, ON, N2L 3G1, Canada}
\author{Michael J. Ditty}
\affiliation{Institute for Quantum Computing, University of Waterloo, Waterloo, ON, N2L 3G1, Canada}
\author{Raymond Laflamme}
\affiliation{Institute for Quantum Computing, University of Waterloo, Waterloo, ON, N2L 3G1, Canada}
\author{David G. Cory}
\affiliation{Department of Nuclear Science and Engineering, Massachusetts Institute of Technology, Cambridge, Massachusetts 02139, USA}

\date{\today}

\begin{abstract}
We demonstrate coherent control of two logical qubits encoded in a decoherence free subspace (DFS) of four dipolar-coupled protons in an NMR quantum information processor. A pseudo-pure fiducial state is created in the DFS, and a unitary logical qubit entangling operator evolves the system to a logical Bell state.   The four-spin molecule is partially aligned by a liquid crystal solvent, which introduces strong dipolar couplings among the spins.  Although the system Hamiltonian is never fully specified, we demonstrate high fidelity control over the logical degrees of freedom.  In fact, the DFS encoding leads to higher fidelity control than is available in the full four-spin Hilbert space.
\end{abstract}


\maketitle

The successful transition from quantum information processing to quantum computation will require the ability to efficiently control qubits in the presence of noise.  Decoherence free subspaces (DFSs) are some of the most efficient schemes of avoiding decoherence from noise sources with underlying symmetries \cite{Duan1997,Zanardi1997,Lidar1998}.  There have been many experimental demonstrations of coherent control in DFSs including demonstrations of multiple qubit control \cite{Kwiat2000,Kielpinski2001,Fortunato2002,Mohseni2003,Ollerenshaw2003,Fortunato2003,Haffner2005,Langer2005,Hodges2007,Cappellaro2007}.  However, in most examples the encoding did not protect against the physically relevant noise sources, and the available control elements did not directly match the generators of logical control.  Here we explore a DFS encoding of strongly dipolar-coupled spins where the logical encoding quite naturally fits the internal Hamiltonian structure.  This work builds on extensive theoretical investigations of DFS encoding for systems with time-dependent exchange couplings \cite{DiVincenzo2000,Bacon2000,Wu2002,Zanardi2004}.  To achieve a similar internal Hamiltonian structure we employ liquid crystal solvents to partially align the spin system and to reintroduce the spin-spin dipolar coupling \cite{EmsleyAndLindon}.

Liquid crystals in a strong external magnetic field are partially ordered.  This partial ordering restricts the thermal motion of molecules dissolved in the liquid crystal material, and consequently, the solute molecules have a preferred orientation, and the orientationally-dependent intramolecular dipolar interactions do not average to zero \cite{EmsleyAndLindon}.  However the translational motion of solute molecules is not restricted, and intermolecular dipolar couplings do average to zero.  The resulting system Hamiltonian for a liquid crystal solvent system is 
\begin{align}
\label{eq:LXhamilt}
\mathscr{H} = \sum_j\pi\nu_j\sigma^j_z+&
					\sum_{j<k}\frac{\pi}{2}\left( J_{jk}+2d_{jk} \right)\sigma_z^j\sigma_z^k + \nonumber \\
				&	\sum_{j<k}\frac{\pi}{2}\left( J_{jk}- d_{jk} \right)\left(\sigma_x^j\sigma_x^k+\sigma_y^j\sigma_y^k\right)
\end{align}
where $\nu_j$ is the resonance frequency of the $j^{\text{th}}$ spin, $d_{jk}$ is the dipolar coupling strength between spins $j$ and $k$, $J_{jk}$ is the corresponding scalar coupling strength, and the sums are restricted to spins within the molecule. In liquid crystal solvent systems, dipolar coupling strengths can reach several kHz, and for a given pair of spins the dipolar coupling is typically one to two orders of magnitude larger than the scalar coupling.  The resonance frequencies and scalar couplings can be directly measured using multiple pulse sequences that average out the dipolar interaction, such as the MREV-8 sequence \cite{Mansfield1971, Mansfield1973, Rhim1973}.  The intramolecular dipolar coupling strengths for a partially ordered system are a result of the structure of the molecule as modified by the order parameters of the system \cite{EmsleyAndLindon,DiehlText}.  Quantum information processing has previously been explored in partially-oriented dipolar-coupled systems \cite{Yannoni1999,Marjanska2000,Mahesh2002}, with as many as seven spins \cite{Lee2006}, and strongly modulating radio frequency (rf) control sequences \cite{Mahesh2006}.

We are interested in the control of two logical qubits that are encoded to protect against collective $\sigma_z$ noise.  The logical subspace $\Sp_L$ for this encoding is the zero-quantum subspace of the Zeeman energy eigenstates
\begin{eqnarray}
\label{eq:logicalstates}
\DvR{00}_L =& \DvR{0101} \\
\DvR{01}_L =& \DvR{0110} \\
\DvR{10}_L =& \DvR{1001} \\
\DvR{11}_L =& \DvR{1010}.
\end{eqnarray}
We describe the system in terms of the set of logical subspace Pauli operators
\begin{align}
\label{eq:logicalops}
\sigma^{L1}_x =& \frac{1}{2}(\sigma^1_x\sigma^2_x+\sigma^1_y\sigma^2_y) & 
		\sigma^{L2}_x =& \frac{1}{2}(\sigma^3_x\sigma^4_x+\sigma^3_y\sigma^4_y) \nonumber \\
\sigma^{L1}_y =& \frac{1}{2}(\sigma^1_y\sigma^2_x-\sigma^1_x\sigma^2_y) & 
		\sigma^{L2}_y =& \frac{1}{2}(\sigma^3_y\sigma^4_x-\sigma^3_x\sigma^4_y) \nonumber \\
\sigma^{L1}_z =& \sigma^1_z-\sigma^2_z & 
		\sigma^{L2}_z =& \sigma^3_z-\sigma^4_z  \nonumber \\
\ident^{L1} =& \frac{1}{2}(\ident^{1,2}-\sigma^1_z\sigma^2_z) & 
		\ident^{L2} =& \frac{1}{2}(\ident^{3,4}-\sigma^3_z\sigma^4_z) \nonumber
\end{align}
along with the nine bipartite terms such as $\sigma^{L1}_x\sigma^{L2}_y$.  Recently we reported on liquid state NMR experiments that demonstrate coherent control for a Bell state with this encoding \cite{Hodges2007}.  We discussed leakage out of the logical subspace under the control operations \cite{Cappellaro2006}, and we described a convenient subsystem pseudo-pure state \cite{Cappellaro2007}.  Here we use a liquid crystal system to extend these studies, introducing a new symmetry for the spin system Hamiltonian that leads us to expect that the logical encoding will be a more natural and efficient subspace for manipulating quantum information.  In particular, the dipolar Hamiltonian has a portion that transforms as the exchange operator which has been shown to be convenient for subsystem encodings \cite{DiVincenzo2000,Bacon2000,Wu2002,Zanardi2004}. 

\begin{figure}
\begin{center}
  \includegraphics[width=3.2in]{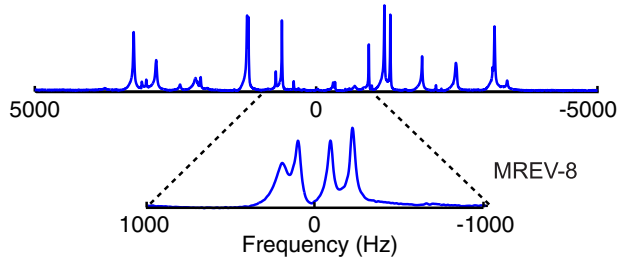}
	\end{center}
	\caption{600 MHz proton spectrum of o-chloronitrobenzene (CNB) partially oriented by the liquid crystal solvent ZLI-1132, where the signal due to solvent protons has been suppressed by inserting a single cycle of the C-48 pulse sequence \cite{Cory1990} ($\tau=30\mu$s) before the acquisition. The inset spectrum, collected stroboscopically under the MREV-8 sequence ($\tau=15\mu$s), shows the four spins uncoupled with chemical shifts scaled by approximately 0.50.}
	\label{ch5fig:CNBspectrum}
\end{figure}

The goal of this work is to demonstrate three results: (1) improved quantum information processing by using logical qubits, (2) an implementation of a DFS with dipolar-coupled spins, and (3) high fidelity control even when we have limited knowledge of the system Hamiltonian.  The spin system used in these studies consists of the four protons of o-chloronitrobenzene (CNB) dissolved in Merck ZLI-1132 liquid crystal at 14 T field and a temperature of 300 K.   The proton spins are strongly coupled to each other, and all of the resonances are not resolved in the 1-D NMR spectrum shown in Figure \ref{ch5fig:CNBspectrum}.
The order parameters for CNB aligned in a liquid crystal solvent have not been measured previously so we do not have this information on which to determine the internal Hamiltonian.  We have made the following simple measurements to obtain reasonable estimates of the internal Hamiltonian: (1) a 1-D MREV-8 spectrum shown in Figure \ref{ch5fig:CNBspectrum}, and (2) 2-D correlation spectra between the chemical shifts under MREV-8 line narrowing and the full internal Hamiltonian (not shown).  The MREV-8 spectrum indicates the chemical shifts, and the 2-D measurement provides a means of assigning the largest dipolar couplings to the appropriate chemical shifts.  We measured the chemical shifts (in units of Hz) $\nu_1=115$, $\nu_2=-234$, $\nu_3=204$, and $\nu_4=-86$ relative to an arbitrary transmitter frequency, and an incomplete set of approximate dipolar couplings (in units of Hz) $d_{12}=-729$, $d_{23}=-503$, and $d_{34}=-1875$.  Although this limited description gives a very incomplete picture of the total system dynamics, it is sufficient for our purposes.  

We encoded two logical qubits into the four spin system where $d_{12}$ provided the control elements to rotate the first logical qubit, $d_{34}$ similarly controlled the second logical qubit and $d_{23}$ controlled the interactions between qubits.  To achieve the desired control fidelity we found pulse sequences via the GRAPE (Gradient Ascent Pulse Engineering) algorithm \cite{Khaneja2005}.  Starting
with an initial guess of the control amplitudes for a pulse of  specified length and number of intervals, the algorithm calculates the forward propagated states or unitaries at every time step, as well as the reverse propagated states or unitaries (from the desired target).  It then uses the local gradient of the overlap between these forward and backward propagated states or unitaries, with respect to the control amplitudes to update the controls.  
Pulses were optimized for robustness to the unknown dipolar couplings ($d_{13}$, $d_{14}$, $d_{24}$), uncertainties ($\pm$100 Hz) in the specified dipolar couplings ($d_{12}$, $d_{23}$, $d_{34}$) and rf inhomogeneity.  The three unknown dipolar couplings were set to a distribution of values ($\pm$100 Hz) centered about a ``best guess'' (in units of Hz) $d_{13}=116$, $d_{14}=-64$, and $d_{24}=-170$.  All scalar couplings are small and were set to zero for pulse optimization.  Although the most accurately known parameters in the internal Hamiltonian are the chemical shifts, control of the logical qubits did not rely on them \cite{CSnote}.  

The experimental goal was to create a pseudo-pure state over the logical qubits and then to entangle them in the form of a Bell state.  We directly created the pseudo-pure state over the logical qubits via temporal averaging.  This was accomplished in two steps.  Under MREV-8 decoupling we prepared the states 
\begin{align}
\sigma_z^{L1}+\sigma_z^{L2}=&\sigma_z^1-\sigma_z^2+\sigma_z^3-\sigma_z^4 \\
\sigma_z^{L2}=&\sigma_z^3-\sigma_z^4
\end{align}
relying on the differences in the chemical shifts of the four spins.  To complete the pseudo-pure state preparation we used a GRAPE pulse to implement 
\begin{equation}
\sigma_z^{L2}\stackrel{U_{prep}}{\longrightarrow}\sigma_z^{L1}\sigma_z^{L2}.
\end{equation}

\begin{figure*}
\begin{center}
  \includegraphics[width=6in]{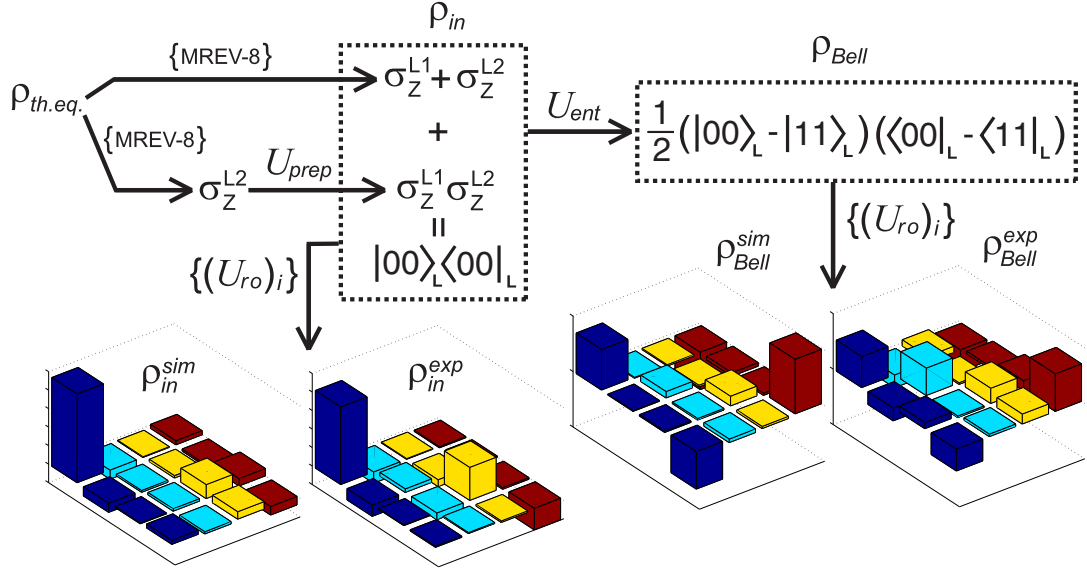}
	\end{center}
	\caption{The procedure and experimental results of creating a Bell state over two logical qubits encoded in the four dipolar-coupled protons of the CNB molecule. Beginning with the thermal equilibrium state, MREV-8 sequences along with a numerically optimized preparation pulse were used to create the pseudo-pure input state $\DvR{00}_L$ over the logical subspace.   A numerically optimized entangling operation pulse converted the input state to the logical Bell state  $(\DvR{00}_L-\DvR{11}_L)/\sqrt{2}$.  A set of 14 numerically optimized readout pulses were used to reconstruct the density matrices shown above.  The experimentally measured logical input and Bell states have correlations of 0.90 and 0.84 (respectively) with the numerically simulated states shown above. }
	\label{ch5fig:flowdiagram}
\end{figure*}

We then implemented
\begin{equation}
\label{eq:Uideal}
U_{ent}^L=\frac{1}{\sqrt{2}}
\left(
\begin{array}{cccc}
	1 & 0 & -i & 0 \\
	0 & -i & 0 & -1 \\
	0 &  i & 0 & -1 \\
	-1 & 0 & -i & 0 \\
\end{array}\right)
\end{equation}
which takes the input state vector $\DvR{00}_L$ to the logical Bell state $\left(\DvR{00}_L-\DvR{11}_L\right)/\sqrt{2}$.  A GRAPE pulse was found which performs this unitary operation over the logical subspace with a simulated fidelity of 0.99, accounting for coherent errors and the uncertainty in Hamiltonian parameters.  In addition to the preparation and entangling pulses, we found fourteen readout pulses that transform every operator in the logical space into measureable operators, as in \cite{Boulant2002}.  The full experiment is outlined in Fig. \ref{ch5fig:flowdiagram}.

Information stored in the DFS respects a direct sum representation over the full Hilbert space.  Ideal control operations over the logical qubits have the structure $\Sp \equiv \SpL \oplus \SpR$, where $\SpL$ is the logical subspace of interest, and $\SpR$ is the remainder of the space.  We attempted to prepare the initial state $\rho_{in}=\DvR{00}_L\DvL{00}_L \oplus \rho^R$, where $\rho^R$ is arbitrary.  The entangling operation over the full Hilbert space should also respect this symmetry $U_{ent}\equiv U_{ent}^L \oplus U_{ent}^R$ where we have complete flexibility in our choice of $U_{ent}^R$.  The challenge of course is that if the direct sum representation is not maintained, the logical information can be corrupted or can leak out of the logical subspace \cite{Wu2002a,Tian2000,Cappellaro2006}.  

$U_{prep}$ and $U_{ent}$ were optimized over the range of dipolar couplings and rf inhomogeneity described above.  The goodness function maximized by the GRAPE algorithm was the logical subspace fidelity \cite{Hodges2007}, which includes a penalty for pulses that permit leakage.  $U^R_{ent}$ was chosen arbitrarily.  Finally, the readout pulses necessarily operated over the entire Hilbert space.  They were simply designed to efficiently transform selected logical operators  into observables and to be robust over the dispersion of coupling constants and rf inhomogeneity.

The reconstructed density matrices over the logical degrees of freedom are shown in Fig. \ref{ch5fig:flowdiagram}.  As expected, most of the observed errors arise from the initial state preparation and the readout sequences.  The normalized state correlations over the logical subspace \cite{Hodges2007} are:
\begin{align}
 \text{Corr}\left(\rho_{in}^{sim}, \rho_{in}^{exp}\right) &= 0.90 \\
 \text{Corr}\left(\rho_{Bell}^{sim}, \rho_{Bell}^{exp}\right) &= 0.84
\end{align}
The normalized correlations between the experimentally measured states and the ideal states are of course lower since they include more of the errors due to state preparation and readout
\begin{align}
 \text{Corr}\left(\rho_{in}^{ideal}, \rho_{in}^{exp}\right) &= 0.83 \\
 \text{Corr}\left(\rho_{Bell}^{ideal}, \rho_{Bell}^{exp}\right) &= 0.76
\end{align}
 
We tested the robustness of our control over the logical subspace by calculating the average state correlation that would be expected as the dipolar frequencies are varied.  The average correlation between simulation and experiment was found to decrease very little (less than 0.03) even when all six dipolar frequencies were simultaneously varied by $\pm$100 Hz.  This is expected, since the pulse sequences were engineered to be robust to these variations \cite{EPAPSrob}.  This result shows that even in the presence of some uncertainty in our Hamiltonian parameters, we have reliable control over the logical subspace of our four proton system.  Finally we compare our control over the encoded logical qubits to that we have over the individual spins.  In numerical simulations, we compared the logical qubit entangling pulse implemented in the experiment to the best spin qubit entangling pulses found by the same pulse optimization methods, optimizing over the full Hilbert space.   The decoherence rate of the logical qubits is slower than the decoherence rate of the spin qubits due to the noise protection of the DFS.  The spin qubit control sequences must be long enough to resolve the differences in spin resonance frequencies ($|\nu_j-\nu_k|$) and therefore are necessarily longer in time than the logical qubit control sequences.  As a result, when the output states are attenuated by decoherence, the fidelity of entangling logical qubits is approximately 0.90, while the fidelity of entangling spin qubits is significantly lower, approximately 0.70 in the best case \cite{EPAPSfid}.  This result quantifies the improvement in control enabled by our logical qubit encoding.

In conclusion, we have shown that improved coherent control is achieved by encoding logical qubits in our system of four dipolar-coupled protons.  Control in this system was demonstrated experimentally by creating a pseudo-pure state in a DFS and applying a unitary transformation to create a logical Bell state.  Though the inital and final states in our experiment are not pure, the implemented transformation is still entangling.
Some of the system Hamiltonian parameters are imprecisely known, and high fidelity control was achieved by engineering pulse sequences that are robust to these uncertainties.  The dipolar couplings in the system Hamiltonian, in the strong coupling regime, provides a natural setting for experimental studies of logical qubit encodings \cite{Yannoni1999}.  In the future, liquid crystal solvent NMR QIPs could be used to explore more complex logical encodings in larger Hilbert spaces.  In these larger systems we expect that the nearest-neighbor couplings will not dominate to such an extent, and that the system will not map onto a one-dimensional chain, as did this 4-spin example.

The authors thank T. S. Mahesh for helpful advice. This material is based upon work supported under an NSF Graduate Research Fellowship, in addition to support by the National Security Agency (NSA) under Army Research Office (ARO) contract W911NF0510469 and by the Air Force Office of Scientific Research.  
 
\bibliography{LXLogical_resubmit2}

\end{document}